\documentclass{article}
\usepackage{arxiv}
\usepackage{times}
\usepackage{helvet}
\usepackage{courier}
\usepackage{comment}
\usepackage{amsmath}
\usepackage{graphicx}
\usepackage{caption}
\usepackage{subcaption}
\usepackage{algorithm}
\usepackage{xcolor}
\usepackage{amsthm, amssymb}
\usepackage[noend]{algpseudocode}
\def\BState{\State\hskip-\ALG@thistlm}
\usepackage{kbordermatrix}

\usepackage[T1]{fontenc}    
\usepackage{hyperref}       
\usepackage{url}            
\usepackage{booktabs}       
\usepackage{amsfonts}       
\usepackage{nicefrac}       
\usepackage{microtype}      

\DeclareMathOperator*{\argmax}{\arg\;\!\!\max}

\usepackage[english]{babel}

\algnewcommand\algorithmicforeach{\textbf{for each}}
\algdef{S}[FOR]{ForEach}[1]{\algorithmicforeach\ #1\ \algorithmicdo}

\title{Speeding Up Distributed Pseudo-tree Optimization Procedure with Cross Edge Consistency \\ to Solve DCOPs}

\author{
  Mashrur Rashik \\
  Department of Computer Science and Engineering\\
  University of Dhaka\\
  \texttt{mashrur639@gmail.com} \\
   \And
 Md. Musfiqur Rahman \\
  Department of Computer Science and Engineering\\
  University of Dhaka\\
  \texttt{musfiq14shohan@gmail.com} \\
  \And
  Md. Mamun-or-Rashid \\
  Department of Computer Science and Engineering\\
  University of Dhaka\\
  \texttt{mamun@cse.du.ac.bd} \\
  \And
  Md. Mosaddek Khan \\
  Department of Computer Science and Engineering\\
  University of Dhaka\\
  \texttt{mosaddek@du.ac.bd} \\
}

\begin{document}

\maketitle
\maketitle
\begin{abstract}

	Distributed Pseudo-tree Optimization Procedure (DPOP) is a well-known message passing algorithm that has been used to provide optimal solutions of Distributed Constraint Optimization Problems (DCOPs)\:-- a framework that is designed to optimize constraints in cooperative multi-agent systems. The traditional DCOP formulation does not consider those constraints that must be satisfied (also known as hard constraints), rather it concentrates only on soft constraints. However, the presence of both types of constraints are observed in a number of applications, such as Distributed Radio Link Frequency Assignment and Distributed Event Scheduling, etc. Although the combination of these types of constraints is recently incorporated in DPOP to solve DCOPs, scalability remains an issue for them as finding an optimal solution is NP-hard. Additionally, in DPOP, the agents are arranged as a DFS pseudo-tree. Recently it has been observed that  the constructed pseudo-trees in this way often come to be chain-like and greatly impair the algorithm's performance. To address these issues, we develop an algorithm that speeds up the DPOP algorithm by reducing the size of the messages exchanged and increasing parallelism in the pseudo tree. Our empirical evidence suggests that our approach outperforms the state-of-the-art algorithms by a significant margin.


\end{abstract}

\section{Introduction}
	Distributed Constraint Optimization Problems (DCOP) are a framework involving multiple agents that are used to interact with one another to achieve a common goal \cite{yokoo1998distributed}. A number of real world problems, such as distributed event scheduling \cite{maheswaran2004taking}, scheduling smart home devices \cite{fioretto2017multiagent} and allocating tasks in mobile sensor networks \cite{jain2009dcops}, can be modelled with this framework. Specifically, a DCOP consists of a number of distributed cost functions which collectively form a global objective function (i.e. the common goal). Each of these cost functions represents a constraint relationship among a set of variables that are controlled by the agents contributing to that constraint. In more detail, each agent is responsible for setting value(s) of its own variable(s) from a finite domain(s). However, they can communicate with their neighbouring agents, and thus can influence value assignment of each other. The goal of a DCOP solution approach is to set every variable to a value from its domain in order to minimize the number of constraint violations or maximize the global objective function.

	
	Over the last couple of decades, a number of algorithms have been proposed to solve DCOPs, and they have been primarily classified into two types: incomplete and complete algorithms. The former experiences better computation and communication cost at the expense of solution quality. Among the incomplete DCOP algorithms DBA \cite{hirayama2005distributed}, DSA \cite{zhang2005distributed} and Max-Sum \cite{farinelli2008decentralised} are the most notable ones. Although it is obvious that this class of algorithms perform well in terms of computation and communication cost, a good number of applications, such as Wi-Fi Channel Assignment \cite{orden2018spectrum}, Reactive Network Resilience \cite{de2017distributed} and many other besides, cannot afford sacrificing the quality of solution. In effect, a number of complete DCOP algorithms have been proposed in the literature, and a lot of efforts can be seen to improve those algorithms. This class of algorithms can be further classified as search-based and inference-based algorithms. The former  use a search technique to find the optimal solution from a set of possible assignments. Some of the notable search-based complete algorithms are SyncBB \cite{hirayama1997distributed}, ConcFB \cite{netzer2012concurrent}, ADOPT \cite{modi2005adopt}. On the other hand, the latter, such as DPOP \cite{petcu2005scalable}, Action-GDL \cite{vinyals2009generalizing}, BrC-DPOP \cite{fioretto2014improving}, are based on dynamic programming techniques. Among them, Distributed Pseudo-tree Optimization Procedure (DPOP) has gained particular attention from the DCOP community. This is due to the fact that DPOP requires a linear number of messages compared to the search-based complete algorithms.
	

	To date, several DPOP variants have been proposed. Specifically, O-DPOP \cite{petcu2006odpop} and MB-DPOP \cite{petcu2007mb} have made improvements in terms of memory requirements of the original algorithm. Then an extension of DPOP, named SS-DPOP \cite{greenstadt2007ssdpop}, improves participating agents' privacy. Whereas, a partially centralized version of DPOP (i.e. PC-DPOP) achieves shorter runtime but sacrifices some privacy \cite{petcu2007pc}. A notable issue with all of the above variants is that they are not able to handle such constraints that must be satisfied (i.e. hard constraints). In contrast, a soft constraint poses a profit/loss for each possible value assignment to its corresponding variables. Nonetheless, the hard constraints are frequently seen in a number of well-known DCOPs, such as distributed Radio Link Frequency Assignment (RLFA)\cite{cabon1999radio} and distributed event scheduling problem\cite{maheswaran2004taking}. In the wake of this shortcoming, two notable extensions of DPOP, H-DPOP \cite{kumar2008h} and BrC-DPOP \cite{fioretto2014improving}, have been proposed. 
	
	In more detail, H-DPOP reduces the computation cost of DPOP by ruling out infeasible combination of the variables, and thus generating smaller messages. This is done by a Constraint Decision Diagram (CDD), which graphically represents a solution set for n-ary constraints \cite{cheng2005constrained}. To do so, H-DPOP performs join and projection operations on CDDs that are computationally expensive. At the same time, it is not possible to fully exploit hard constraints to prune the domain of a variable using this approach. This particular issue has been addressed by BrC-DPOP through the use of Value Reachability Matrix (VRM) which is a representation of a constraint between two variables in the form of a matrix. It is worth noting that similar to the aforementioned DPOP extensions, BrC-DPOP uses depth-first search pseudo tree to graphically represent a DCOP. Notably, it is shown in \cite{chen2017improved} that a depth first search pseudo tree often results in a chain-like structure thus impairing the performance of the algorithm due to the lack of parallelism. Nevertheless, the algorithm proposed in the paper, the so-called BFS-DPOP, has shown the significance of an alternative graphical representation$-$ breadth-first search pseudo tree. To be exact, BFS-DPOP enhances parallelism, and thus reduces the runtime of the algorithm. However, BFS-DPOP cannot handle hard constraints, and thus is not directly applicable to BrC-DPOP.

	Against this background, we propose a new variant of the DPOP algorithm, that we call CeC-DPOP. CeC-DPOP takes the advantage of parallelism through the use of BFS pseudo tree as the communication structure. It can also deal with hard constraints. However, unlike BrC-DPOP that enforces branch consistency, CeC-DPOP uses a new form of consistency, namely Cross-edge Consistency (CeC). This particular phenomenon enables CeC-DPOP to produce smaller message size and improve DPOP's runtime of by pruning the domain of the corresponding variables. To be precise, We empirically evaluate the performance of our approach, and observe a significant reduction of runtime, average of 60\% by using this technique.



\section{Background and Problem Formulation}

\begin{figure}[t]
	\centering
	\begin{subfigure}[H]{.3\textwidth}
		\centering
		\includegraphics[scale=0.2]{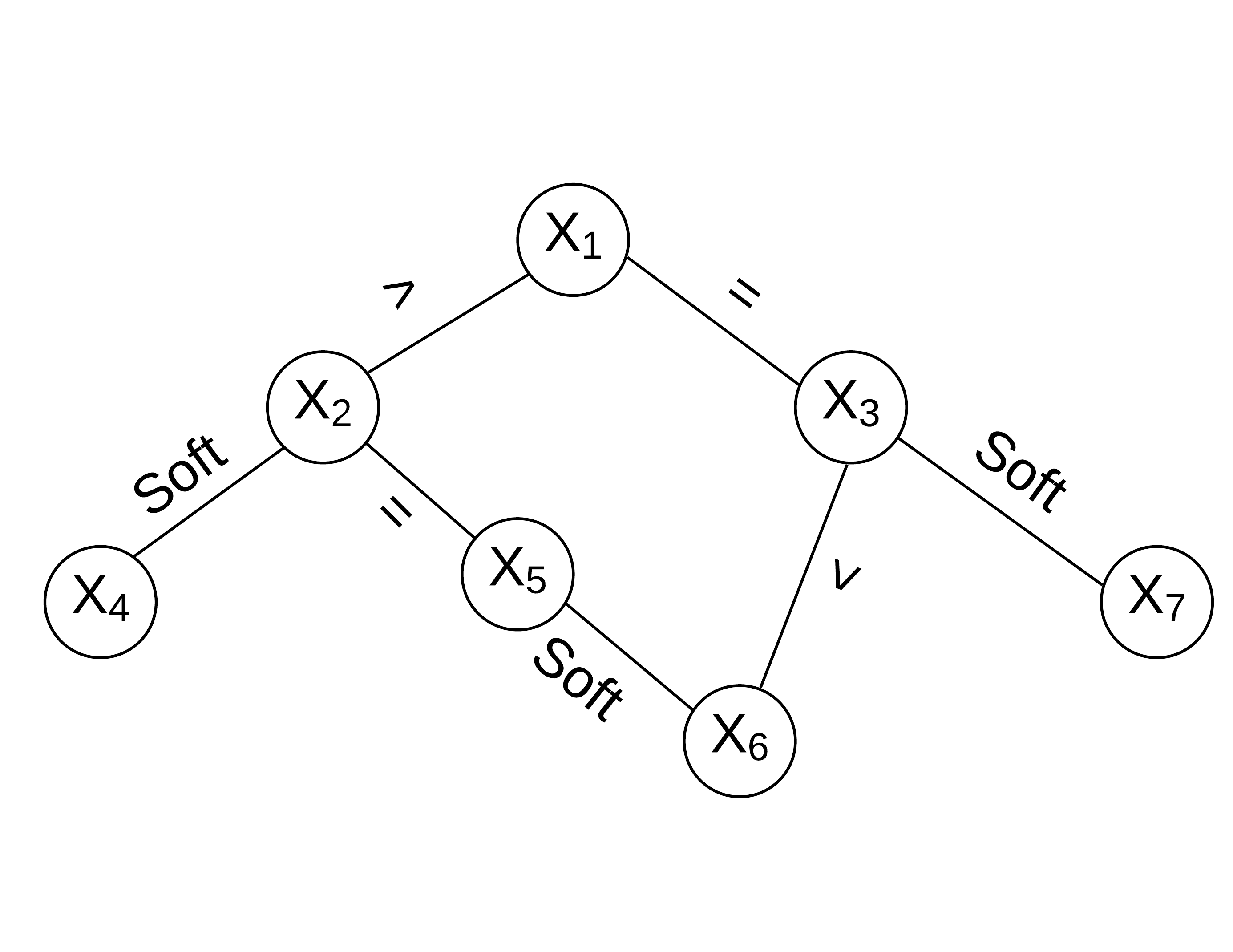}
		\caption{A constraint graph representation of a DCOP. Here, the edges having relational operators are the hard constraints.}
		\label{fig:constraintgraph}
	\end{subfigure}
	\hfill
	\begin{subfigure}[H]{.5\textwidth}
		\centering
		\begin{table}[H]
			\centering
			\begin{tabular}{|c|c|c|}
					\hline
					$x_5$                       & $x_6$                      & Cost                  \\ \hline
					0                        & 0                       & 12                    \\ \hline
					0                        & 1                       & 3                     \\ \hline
					1                        & 0                       & 7                     \\ \hline
					1                        & 1                       & 3                     \\ \hline
			\end{tabular}
		\end{table}
		\caption{A sample cost table for soft constraint involving variable $x_5$ and $x_6$. }
		\label{fig:utilitytable}
	\end{subfigure}
	\caption{Example of a DCOP Instance containing both soft and hard constraints}
\end{figure}
\setlength{\textfloatsep}{2mm}
\setlength{\floatsep}{2mm}

\par A DCOP model can be formally expressed as a tuple $\langle$\textbf{A}, \textbf{X}, \textbf{D}, \textbf{F}, $\alpha\rangle$ where:

\begin{itemize}
	\item \textbf{A} = \{$a_1,a_2,....,a_k$\} is a set of agents.
	\item \textbf{X} = \{$x_1,x_2,....,x_n$\} is a set of variables, where n$\geq$k.
	\item \textbf{D} = \{$d_1,d_2,....,d_n$\} is a set of domains for the variables in \textbf{X}, where $d_i \in D$ is the available domain for the corresponding variable $x_i \in X$.
	
	\item \textbf{F} = \{$f_1,f_2,....,f_m$\} is a set of constraint functions (also known as utility or cost functions). Here, each function $f_i( \mathbf{x_i})$ depends on a subset of variables $\mathbf{x_i} \subseteq $ \textbf{X} that can be mentioned as the scope of that function. In order to represent the relationship among the variables in $\mathbf{x_i}$, the function $f_i( \mathbf{x_i})$ denotes the utility value for each possible assignment of those variables. Each constraint $f_i \in$  \textbf{F} can be hard in which the value combinations that must be avoided are denoted as the cost 0 and the combinations that are allowed have the cost 1. The remaining type is the soft constraint indicating that each value combination results in a finite utility/cost value and need not to be avoided. The dependencies among the variables can be used to construct a constraint graph that has been used to represent DCOPs graphically. In this representation, each variable is associated with a node and connected to each other through an edge.
	
	\item $\alpha$ : \textbf{X} $\rightarrow$ \textbf{A} is an onto mapping function that assigns the variables \textbf{X} to the set of agents \textbf{A}.
	 
\end{itemize}
\begin{equation}
	\mathbf{X^*} = \argmax_x \sum_{f_i\in F} f_i(x_i)
\end{equation}
Within this model, the main objective of a DCOP algorithm can be expressed as each agent assigning the values to its associated variable(s) from the corresponding domain(s) that can be expressed as $\textbf{X}^*$, in the pursuit of maximization or minimization of the sum of the utility functions (i.e. the global objective function). In this paper, we are going to consider the maximization problem only (Equation 1).
For example, in Figure~\ref{fig:constraintgraph} ,a DCOP instance is graphically represented as a constraint graph. Here, we consider the set of variables $\textbf{X}=\{x_1,x_2,...,x_7\}$,
each having domain $d_i=\{0,1\}$. The cost matrix of the soft constraint involving variables $x_5$ and $x_6$ is showed in figure ~\ref{fig:utilitytable}. The remaining constraints in the graph that are defined by relational operators are hard constraints.

\par
\begin{figure}[t]
    	\centering
    	\begin{subfigure}[H]{.4\textwidth}
    		\centering
    		\includegraphics[scale=0.2]{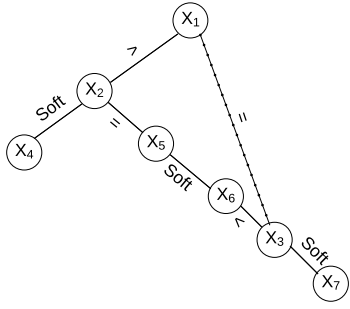}
    		    		\caption{DFS Pseudo-tree}
    		\label{fig:dfsgraph}
    	\end{subfigure}
    		\hfill
    	\begin{subfigure}[H]{.4\textwidth}
    		\centering
    		\includegraphics[scale=0.2]{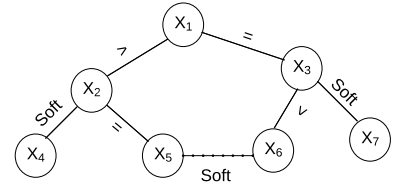}
    		
    		\caption{BFS Pseudo-tree}
    		\label{fig:bfsgraph}
    	\end{subfigure}
    		\vspace{-3mm}
    		\caption{DCOP Pseudo-trees}
    \end{figure}

 As aforementioned, Distributed Pseudo-tree Optimization Procedure is a complete, synchronous message passing algorithm for solving DCOPs. Specially it uses the dynamic programming technique on a DFS pseudo-tree in a distributed manner. DPOP is executed through three phases.
	 In the first phase, a distributed DFS traversal is started from the root(held by an agent) of the constraint graph using the distributed DFS algorithm like in \cite{petcu2008m}.
	  As a result, a DFS pseudo-tree structure is built where each agent labels its neighbours as parents, pseudo-parents, children or pseudo-children and edges are identified as tree or back edges. 
	  For example, after this phase, the constraint graph in Figure ~\ref{fig:constraintgraph} will result in a DFS pseudotree like in Figure ~\ref{fig:dfsgraph}. The resulting DFS pseudo-tree serves as a communication structure for the subsequent phases of DPOP. The next phase is the Util propagation phase in which each agent, starting from the leaves of the constraint graph, sends UTIL message to its parent. The UTIL message is generated by aggregating the constraint utilities between the current node and the variables in its separator that is the ancestors of the current node that are connected directly to this node or its descendants and also the utilities in the UTIL message received from its children and finally projecting out its own variables by optimizing over them. At last, the value propagation phase is initiated from the root agent. Each agent selects its optimal assignment using the cost function computed in the UTIL propagation phase and the VALUE message received from its parent. Afterwards, each agent broadcasts its assignment to its children. When every agent has chosen its optimal assignment, the algorithm terminates. 

	\par
	DPOP can be executed on different branches independently using DFS pseudo-tree as communication structure. Though DPOP produces linear number of messages as mentioned before, message size in  this algorithm is exponential. Another notable limitation of the DPOP algorithm is that it does not exploit hard constraint along with soft constraints which has been found useful in many real life DCOP problems. These two situation can be resolved by another algorithm BrC-DPOP proposed by \cite{fioretto2014improving}.

    \par BrC-DPOP exploits hard constraints by enforcing arc consistency and introducing a weaker form of the path consistency which can be applied along the path of a pseudo-tree in pursuit of reducing message size.The algorithm starts with generating a pseudo-tree structure followed by a path construction Phase which is later used to get the knowledge of the direct paths from each agent to its parent and pseudo-parents. In the next phase, arc consistency is enforced in a distributed environment.Then the most important phase is executed where branch consistency is exploited in a distributed way. The aim of this phase is to ensure mutual reachability of every pair of values of an agent and its pseudo-parents considering every pseudo-tree path between them. Finally, the UTIL and VALUE propagation phase are executed considering the updates of the pseudo-tree. The advantages of BrC-DPOP includes smaller message size due to BrC propagation enforcement as well as faster runtime since it prunes the values of the variables. Though BrC-DPOP improves the DPOP algorithm to a great scale, the communication structure is DFS pseudo-tree, and as previously mentioned often becomes chain-like in many experiments for example in Figure~\ref{fig:dfsgraph}. This condition greatly reduces performances of solving DCOPs by Brc-DPOP or other variants of DPOP that use DFS pseudo-tree as the communication structure. To deal with this drawback, \cite{chen2017improved} propose a variant of DPOP algorithm (the so-called BFS-DPOP) which uses the Breadth First Search (BFS) pseudo-tree as the communication structure.

    \par In more detail, BFS-DPOP operates on Breadth First Search (BFS) pseudo-tree that is used as the
    communication structure intending to increase parallelism. This is because it produces more branches than that of the DFS counterpart. Here, Figure~\ref{fig:bfsgraph} depicts the transformed BFS pseudo-tree of the corresponding constraint graph of Figure ~\ref{fig:constraintgraph}. In BFS-DPOP, following the construction of BFS Pseudo-tree, the cluster removal phase occurs wherein the allocation of cross-edges are decided so that it can reduce the maximal message size as much as possible by the disposal of cross-edge constraints. Finally, the UTIL and VALUE propagation phase is executed on the BFS Pseudo-tree considering the changes occurred in the previous phases. Even though BFS-DPOP experiences shorter communication paths, and hence less communication time, through the use of BFS pseudo-tree, the algorithm produces messages with exponential size as the system grows, as in traditional DPOP. Moreover, as aforementioned, this algorithm can not deal with hard constraints which is utilized in BrC-DPOP by enforcing branch consistency. On the other hand, BFS-DPOP is not suitable for exploiting branch consistency. In light of the above background, we address these issues in the section that follows. 
    

\section{Cross-Edge Consistent DPOP (CeC-DPOP)}
	CeC-DPOP improves DPOP by enforcing cross-edge consistency that reduces the domain size of the variables of a DCOP. In effect, it reduces the message size and runtime of the DPOP algorithm. Moreover, unlike the traditional DPOP algorithms, CeC-DPOP uses BFS pseudo tree instead of a DFS pseudo tree in order to take its inherent benefits of increased parallelism and shorter tree depth. Specifically, this algorithm comprises of four phases, BFS pseudo tree construction, consistency enforcement, UTIL propagation and VALUE propagation phase.

	\begin{algorithm}[t]
		\caption{Path-Construction($G_{bfs}$, \textbf{P}, \textbf{C})}\label{path-construction}
		\textbf{Input:} Pseudo tree $G_{bfs}$, set of parents \textbf{P}, set of child \textbf{C}.\\
		\textbf{Output:} A list containing path information from current variable to enforce cross-edge consistency.
		\begin{algorithmic}[1]
			\ForEach{cross edge $(x_i, x_j)$ of $G_{bfs}$}
			\State $x\textsubscript{\textit{l}}$ $\gets$ $LCA(x\textsubscript{i}, x\textsubscript{j})$
			\State send $NEXT\_UPDATE(x\textsubscript{\textit{l}}, x\textsubscript{i})$ to $P\textsubscript{i}$
			\EndFor
			\If{$x_i$ is not an end point of a cross edge}
			\State send $NEXT\_UPDATE(NULL, x\textsubscript{i})$ to $P\textsubscript{i}$
			\EndIf
			\While{$ cnt\_next\textsubscript{i} < |C\textsubscript{i}| $}
			\If{receive $NEXT(x\textsubscript{\textit{l}}, x\textsubscript{c})$ from $x\textsubscript{c}\in C\textsubscript{i}$}
			\State $NEXT\textsubscript{i}$ $\gets$ $NEXT\textsubscript{i}$ $\cup$ $(x\textsubscript{\textit{l}}, x\textsubscript{c})$
			\EndIf
			\If{receive $complete(x\textsubscript{c})$ from $x\textsubscript{c}\in C\textsubscript{i}$}
			\State $cnt\_next\textsubscript{i}$ $\gets$ $cnt\_next\textsubscript{i} + 1$
			\EndIf 
			\EndWhile
			\If{$NEXT\textsubscript{i}$ not equals \textit{NULL}}
			\ForEach{$x\textsubscript{\textit{l}}$ such that $(x\textsubscript{\textit{l}}, x\textsubscript{c}) \in NEXT\textsubscript{i}$}
			\State send $NEXT\_UPDATE(x\textsubscript{\textit{l}}, x\textsubscript{i})$ to $P\textsubscript{i}$
			\EndFor 
			\EndIf
			\State send $complete(x\textsubscript{i})$ to $P\textsubscript{i}$
		\end{algorithmic}
	\end{algorithm}
	
	Initially, a pseudo tree is constructed from the constraint graph. In order to generate the corresponding BFS pseudo tree, we use the same method as prescribed in the BFS-DPOP algorithm. For example, Figure~\ref{fig:bfsgraph} illustrates a sample BFS pseudo tree of the constraint graph depicted in Figure~\ref{fig:constraintgraph}. For simplicity, we use a part (Figure~\ref{fig:tracegraph}) of the pseudo tree of Figure~\ref{fig:bfsgraph} as the worked example of our algorithm. Having a BFS pseudo tree $G_{bfs}$ constructed, CeC-DPOP enforces arc-consistency. This phase uses the distributed Arc-Consistency (AC) algorithm that is introduced in BrC-DPOP algorithm. This algorithm results in a reduced domain for all the variables having hard constraints, as shown in Figure~\ref{fig:tracearcconsistency}, where domain of each variable is \{0, 1, 2, 3, 4\}. 
	
	After arc-consistency is enforced, CeC-DPOP enforces a new form of consistency (i.e. the so-called cross edge consistency) on the BFS pseudo tree. To do so, we need the lowest common ancestor $LCA(x_i, x_j)$ for every pair of variables $x_i$ and $x_j$ in $G_{bfs}$. To find the LCA of every pair of variables, we followed the distributed algorithm shown in \cite{schieber1988finding}. In order to represent hard constraints we use consistency matrices, where a matrix $M_{ij}$ represents a hard constraint between variables, $x_i$ and $x_j$. The consistency matrix between $x_1$ and $x_2$, which represents the constraint $x_1<x_2$ as shown in Figure 3.  
	
	\vspace{-3mm}
	\begin{figure}[H]
		\centering
		\renewcommand{\kbldelim}{(}
		\renewcommand{\kbrdelim}{)}
		\[
		\text{M}_{12} = \kbordermatrix{
			& 0 & 1 & 2 & 3 & 4 \\
			0 & 0 & 1 & 1 & 1 & 1 \\
			1 & 0 & 0 & 1 & 1 & 1 \\
			2 & 0 & 0 & 0 & 1 & 1 \\
			3 & 0 & 0 & 0 & 0 & 1 \\
			4 & 0 & 0 & 0 & 0 & 0
		}
		\]
		\label{fig:consistencymatrix}
		\vspace{-2mm}
		\caption{An example consistency matrix ($M_{12}$) between $x_1$ and $x_2$ where a 1 along a particular row ($r_i$) and column ($c_i$) indicates that $x_1=r_i$ and $x_2=c_i$ is allowable for $x_1<x_2$}
	\end{figure}
	\vspace{-3mm}
	
	\begin{figure}[t]
		\centering
		\begin{subfigure}[H]{.3\textwidth}
			\centering
			\includegraphics[scale=0.22]{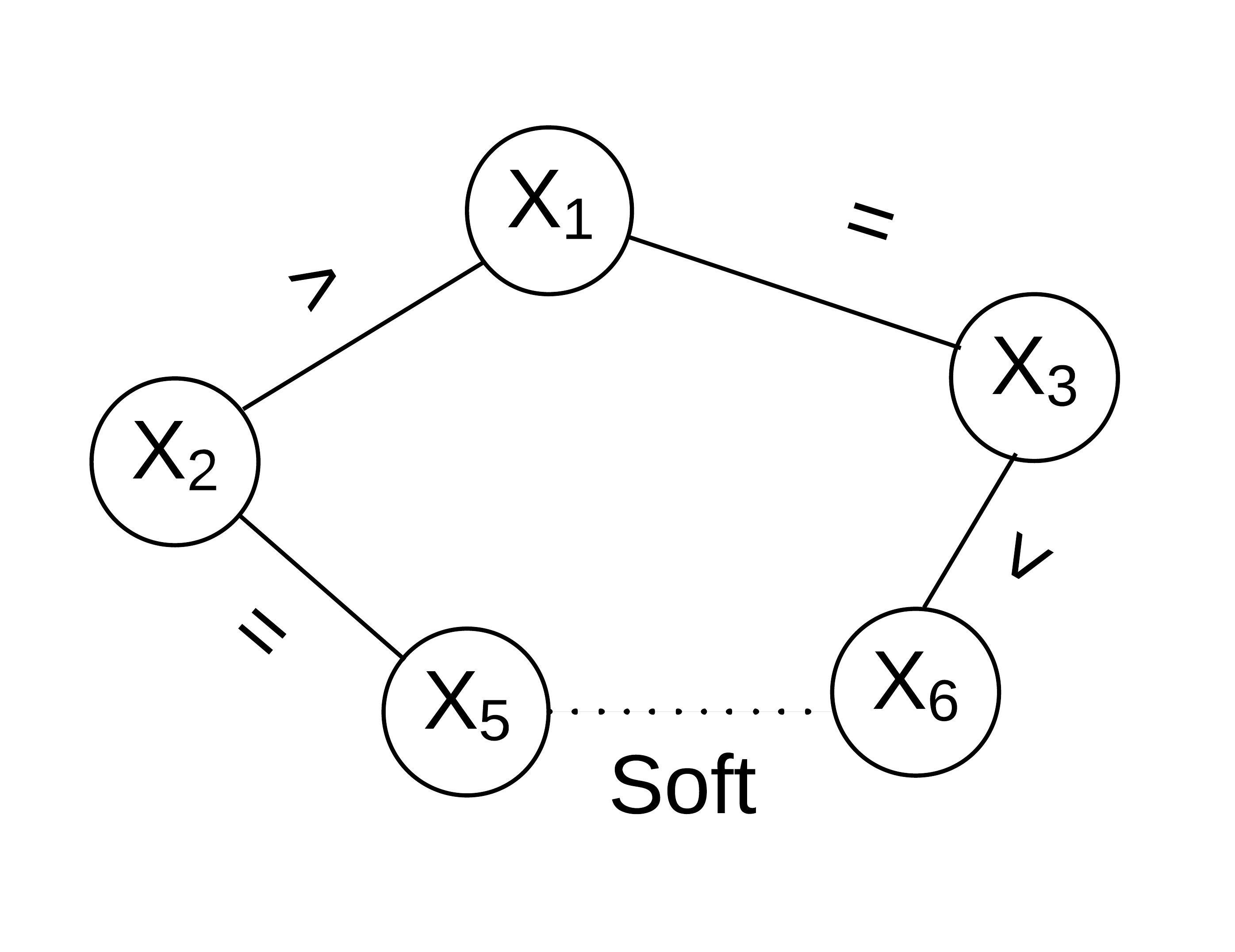}
			\caption{A part of pseudo-tree depicted in Figure~\ref{fig:bfsgraph}, used as an example to enforce cross-edge consistency}
			\label{fig:tracegraph}
		\end{subfigure}
		\hfill
		\begin{subfigure}[H]{.3\textwidth}
			\centering
			\includegraphics[scale=0.25]{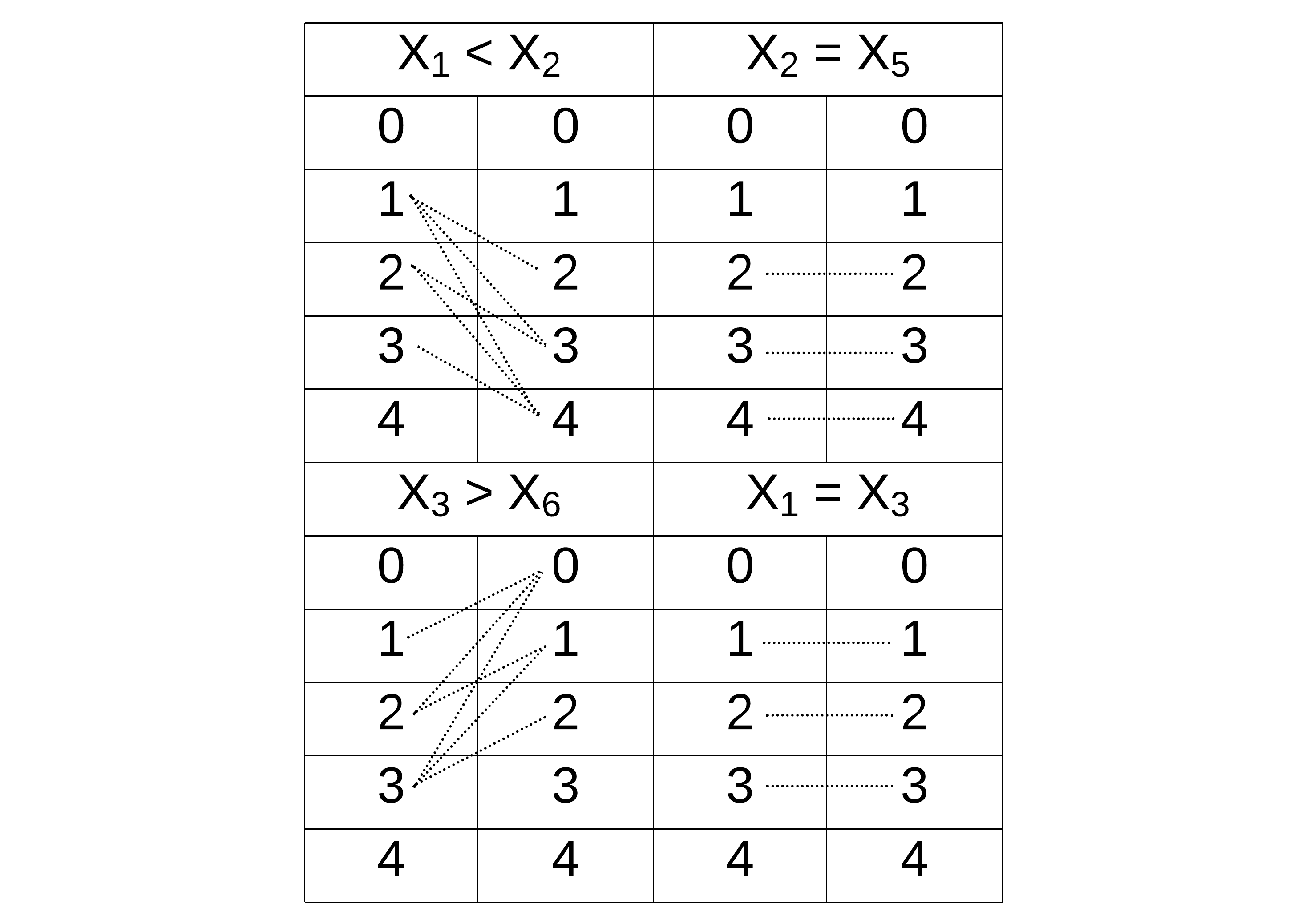}
			\caption{Consistent pairs after arc-consistency enforcement. Here, an arrow indicates an allowable pair}
			\label{fig:tracearcconsistency}
		\end{subfigure} 
		\hfill	
		\begin{subfigure}[H]{.3\textwidth}
			\centering
			\includegraphics[scale=0.25]{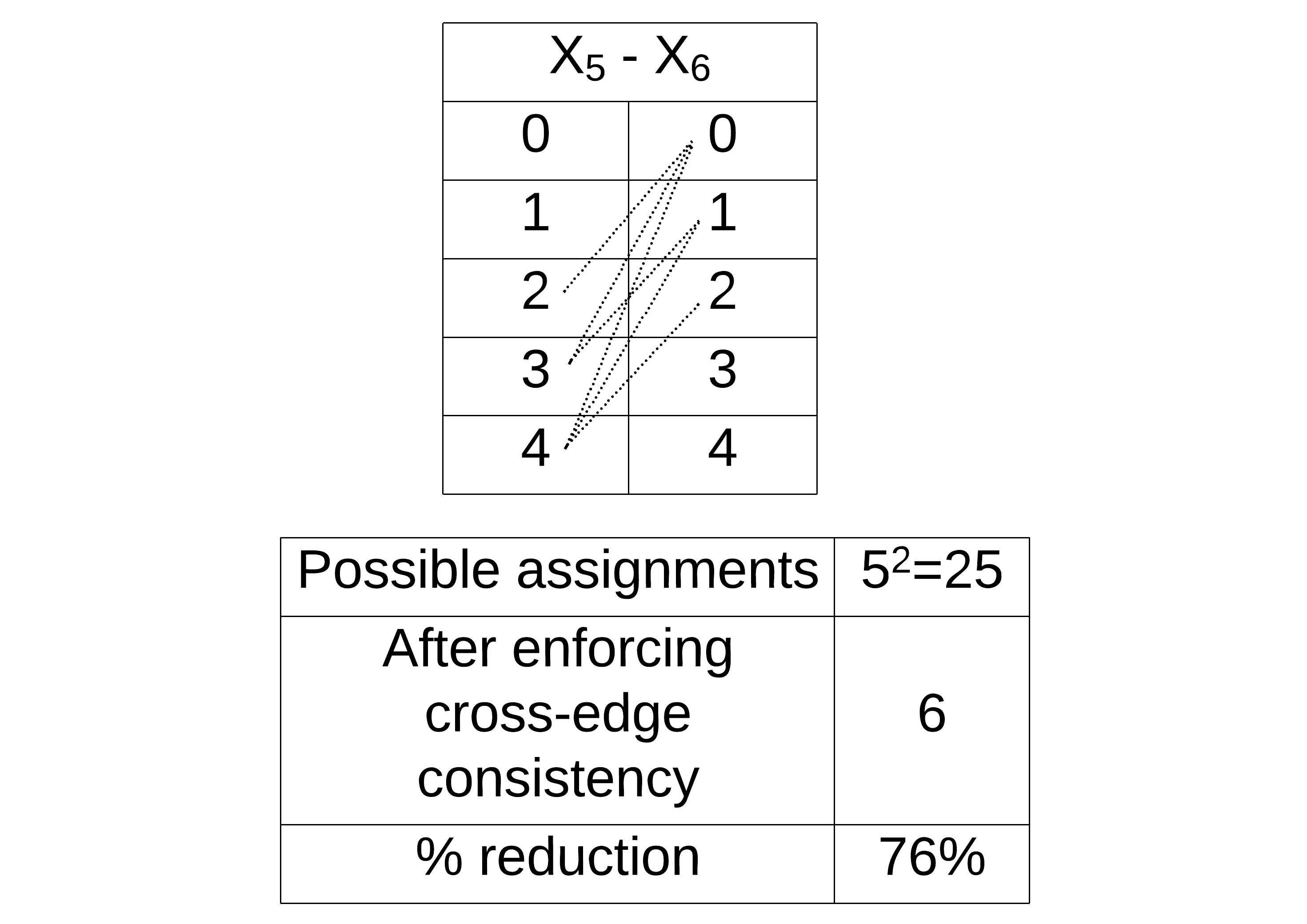}
			\caption{Consistent pairs along cross-edge after applying CeC-DPOP which shows an overall 76\% domain size reduction}
			\label{fig:tracecrossconsistency}
		\end{subfigure}
		
		\caption{Simulation of the algorithm}
	\end{figure}
	
	\begin{algorithm*}[t]
		\caption{Cross-Edge-Consistency-Propagation($G_{bfs}$, \textbf{M}, \textbf{C}, $NEXT$)}\label{ce-consistency}
		\textbf{Input:} Pseudo tree $G_{bfs}$, set of consistency matrix \textbf{M}, set of child \textbf{C}, next list $NEXT$.\\
		\textbf{Output:} A cross edge consistent pseudo tree.
		\begin{algorithmic}[1]
			\If{$x_i$ is root} \hspace{72mm}// CeC message propagation initiated by root
			\ForEach{$x_c$ in $C_i$}
			\State send $CeC(x\textsubscript{i}, M\textsubscript{ii})$ to $x\textsubscript{c}$
			\EndFor 
			\EndIf
			\If{received $CeC(x\textsubscript{p}, M\textsubscript{p\textit{l}})$ from $x\textsubscript{p}$ such that $x_p$ is parent of $x_i$} // propagating the updated consistency matrix
			\ForEach{$(x\textsubscript{\textit{l}}, x\textsubscript{c}) \in NEXT\textsubscript{i}$}
			\If{$x\textsubscript{\textit{l}}$ equals $x\textsubscript{i}$}
			\State $M\textsubscript{i\textit{l}}$ $\gets$ $M\textsubscript{ii}$ \hspace{62mm}// initializing to unary constraint 			
			\Else
			\State $M\textsubscript{i\textit{l}}$ $\gets$ $M\textsubscript{ip}$ $\times$ $M\textsubscript{p\textit{l}}$ 
			\EndIf
			\If{$x\textsubscript{c}$ not equals \textit{NULL}}
			\State send $CeC(x\textsubscript{i}, M\textsubscript{i\textit{l}})$ to $x\textsubscript{c}$ 
			\EndIf
			\EndFor
			\EndIf
			\ForEach{$x\textsubscript{c}\in C\textsubscript{i}$ such that $x\textsubscript{c}$ received no $CeC\textsubscript{i}$ message}
			\State send $CeC(x\textsubscript{i}, \textit{NULL})$ to $x\textsubscript{c}$
			\EndFor
			\ForEach{$ \textit{ce\textsubscript{ij}} \in \textit{CE\textsubscript{i}} $ such that $x\textsubscript{\textit{l}}$ equals $LCA\textsubscript{ij}$} \hspace{23mm}// computing reduced domain along cross-edge
			\State $M\textsubscript{\textit{l}j}$ $\gets$ $M\textsubscript{j\textit{l}}\textsuperscript{T}$ \hspace{68mm} // matrix transpose 
			\State $M\textsubscript{ij}$ $\gets$ $M\textsubscript{i\textit{l}}$ $\times$ $M\textsubscript{\textit{l}j}$
			\EndFor
		\end{algorithmic}
	\end{algorithm*}
	
	Now the algorithm enforces cross edge consistency on the pseudo tree $G_{bfs}$. For this, we need to construct a path for each cross-edge in $G_{bfs}$ (Algorithm~\ref{path-construction}). The BFS pseudo tree $G_{bfs}$, parent set \textbf{P} and set of child \textbf{C} are the inputs of the algorithm. Throughout the algorithm, we construct a $NEXT_i$ list containing information in the form of $(x_l, x_c)$. It informs the current agent $x_i$ about the next agent $x_c$ to enforce cross-edge consistency for an edge whose endpoints have LCA at $x_l$. The for loop in line 1 selects a cross edge having one end point $x_i$ from $G_{bfs}$, and sends a message $NEXT\_UPDATE(x_l, x_i)$ to its parent $P_i$. This message contains information about the LCA $x_l$ of two variables $x_i$ and $x_j$ and the current variable $x_i$. To do this, line 2 computes a LCA, $x_l$ of $x_i$ with another variable $x_j$ with whom it holds cross edge. Then in line 3 $x_i$ sends a $NEXT\_UPDATE(x_l, x_i)$ to its parent $P_i$. In line 4, CeC-DPOP checks whether $x_i$ is a member of a cross edge, and if this is not the case, it sends a $NEXT\_UPDATE(NULL, x_i)$ to parent $P_i$. Here, NULL indicates that $x_i$ is not an end point of any cross edge. In our exemplary pseudo tree of Figure \ref{fig:tracegraph}, $x_5$ computes a LCA (i.e. $x_1$) for its cross edge connecting $x_5$ and $x_6$ and then sends $NEXT\_UPDATE(x_1, x_5)$ to its parent $x_2$.   
	
	Afterwards, the while loop in line 6 compares a counter variable, $cnt\_next_i$ with the child count of current variable (i.e. $|C_i|$) to check whether the current variable received a $NEXT\_UPDATE$ message from each child in $C_i$. Within this loop, if a $NEXT\_UPDATE(x_l, x_c)$ is received from a child, appends $(x_l, x_c)$ to the list, $NEXT_i$ (line 7-8). Then line 9 checks for any $complete(x_c)$ received from a child. This message informs the current variable $x_i$ that path construction for the sub tree rooted at $x_c$ is complete. For each received $complete(x_c)$ message, line 10 increments $cnt\_next_i$ by 1. The while loop terminates when each child $x_c$ in $C_i$ sends a $complete(x_c)$ message. In our example, after receiving $NEXT\_UPDATE(x_1, x_5)$ from $x_5$, $x_2$ appends $(x_1, x_5)$ to $NEXT_2$. Now, in line 11, the algorithm checks whether $NEXT_i$ is not empty. If this is true, the for loop in line 12 selects each $(x_l, x_c)$ pair from the $NEXT_i$ list and line 13 sends a $NEXT\_UPDATE(x_l, x_i)$ message to $P_i$. Next, the algorithm terminates after sending a $complete(x_i)$ message to $P_i$ after line 14. In this context, in our example after $x_5$ sends a next message to $x_2$, $x_5$ generates a $complete(x_5)$ message and sends it to $x_2$. Upon receiving a $complete$ message from its only child $x_2$, increments $cnt\_next_i$ by 1 and the while loop of line 6 terminates. In our example of Figure~\ref{fig:tracegraph}, $x_2$ sends $NEXT\_UPDATE(x_1, x_2)$ to $x_1$. After receiving this, $x_1$ obtains information about its next child $x_2$ to enforce cross edge consistency. At the same time, $x_2$ obtains information about $x_5$. Thus, a path from $x_1$ to $x_5$ through the variable $x_2$ is established. Similarly, another path from $x_1$ to $x_6$ through the variable $x_3$ is established.

	Finally, we enforce cross edge consistency on the path that is established on the pseudo tree (Algorithm \ref{ce-consistency}). The BFS pseudo tree $G_{bfs}$, set of consistency matrices \textbf{M}, set of child \textbf{C} and the $NEXT$ list are the inputs of the algorithm. The algorithm works as follows. Line 1 of the algorithm checks whether the current variable $x_i$ is root. If this true, it initiates CeC message propagation by iterating every child using the for loop in line 2. Line 3 then sends a $CeC(x_i, M_{ii})$ to every $x_c$ in $C_i$, where $x_i$ is the variable which sent the message along with its consistency matrix $M_{ii}$. Line 4 of the algorithm checks whether any CeC message has been received from its parent. If this is the case, line 5 iterates over each pair $(x_l, x_c)$ of the $NEXT_i$ list to propagate CeC message. For this purpose, line 6-7 checks whether the current variable equals to LCA $x_l$ of a cross edge in the subtree. If this is the case, it initializes $M_{il}$ with its unary constraint $M_{ii}$ which represents the domain of the current variable $x_i$. Otherwise, line 8-9 computes $M_{il}$, which is a multiplication of $M_{ip}$ and $M_{pl}$. Next, line 10 checks whether $x_c$ is not null. If this is true, line 11 sends a $CeC(x_i, M_{il})$ message to $x_c$. In our example of Figure \ref{fig:tracegraph}, $x_1$ initializes CeC message propagation and sends a $CeC(x_1, M_{11})$ to both $x_2$ and $x_3$. When $x_2$ receives $CeC(x_1, M_{11})$ from $x_1$, it computes $M_{21}$ by multiplying $M_{21}$ with $M_{11}$ and sends this new consistency matrix to its child $x_5$ through a $CeC(x_2, M_{21})$ message. Now, line 12-13 of the algorithm checks whether any child exists that did not receive any CeC message. If this is true, $x_i$ sends a $CeC(x_i, NULL)$ to that child. Line 14-16 finally computes the consistency matrices along each cross edge by iterating over every cross edge and multiplying the matrices obtained for each end point of the cross edge. In our example, after $x_5$ receives a $CeC(x_2, M_{21})$ from $x_2$, it computes $M_{51}$. In a similar way $x_6$ computes $M_{61}$. To obtain the reduced set of assignable pairs, CeC-DPOP multiplies $M_{51}$ with $M_{16}$. The output for our simulation is shown in Figure \ref{fig:tracecrossconsistency}. Here a total number of allowable pairs is $5^2 = 25$. After cross-edge consistency is enforced, the number of allowable pairs is reduced to 6. Hence, the reduction is $\frac{25-6}{25}=0.76$, which is a 76\% reduction in the total number of assignable pairs along $x_5$ and $x_6$. After cross edge consistency is enforced, we obtain a set of variables with reduced domain size. Now, we execute the UTIL and VALUE propagation phase. These two steps correspond to the UTIL and VALUE propagation phase of BFS-DPOP algorithm.


	

	\begin{figure*}[t]
		\centering
		\begin{subfigure}[H]{.45\textwidth}
			\centering
			\includegraphics[scale=0.45]{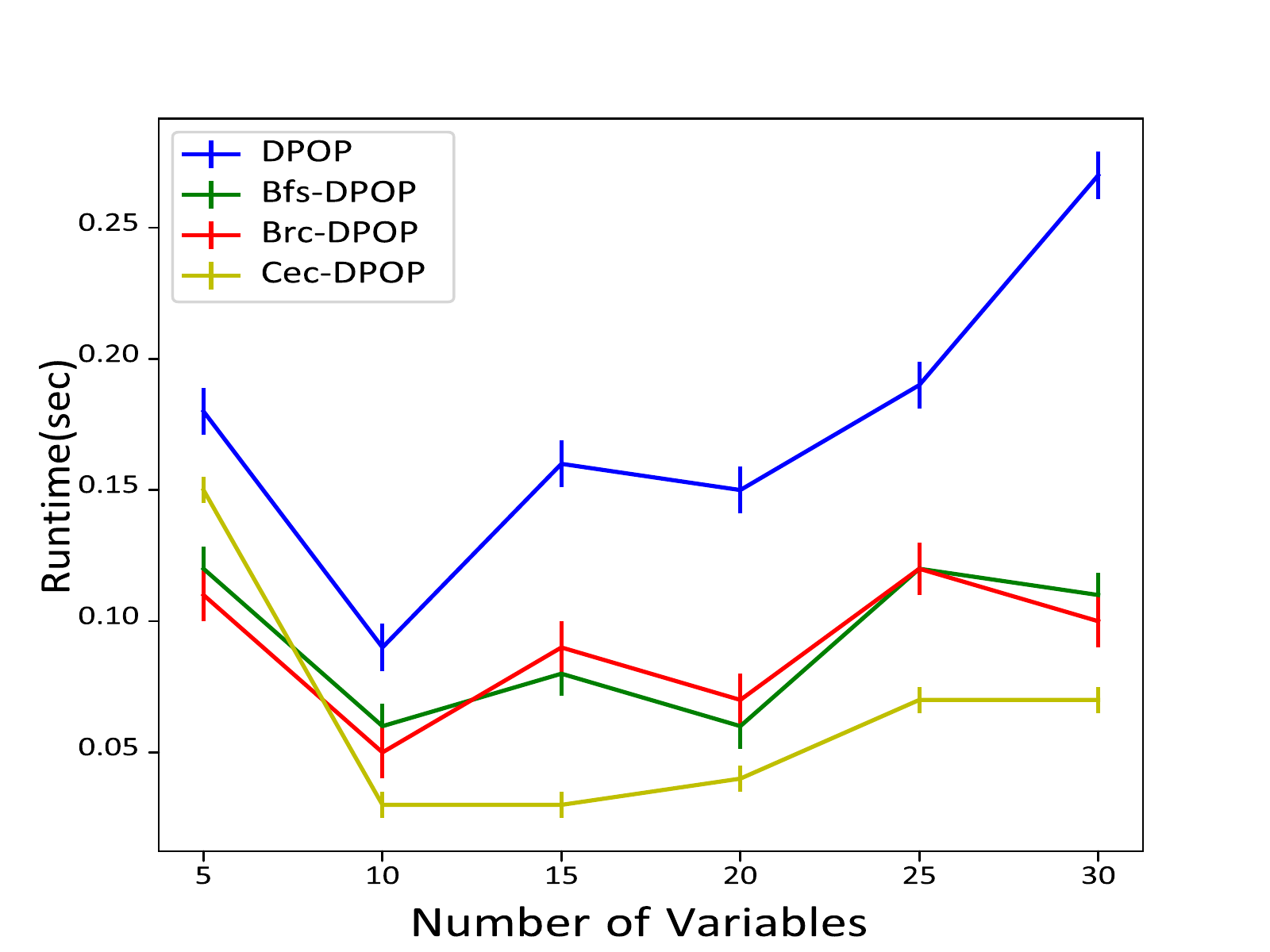}
			\caption{number of variables vs runtime (random graph)}
			\label{graph:one}
		\end{subfigure}
		\hfill
		\begin{subfigure}[H]{.45\textwidth}
			\centering
			\includegraphics[scale=0.45]{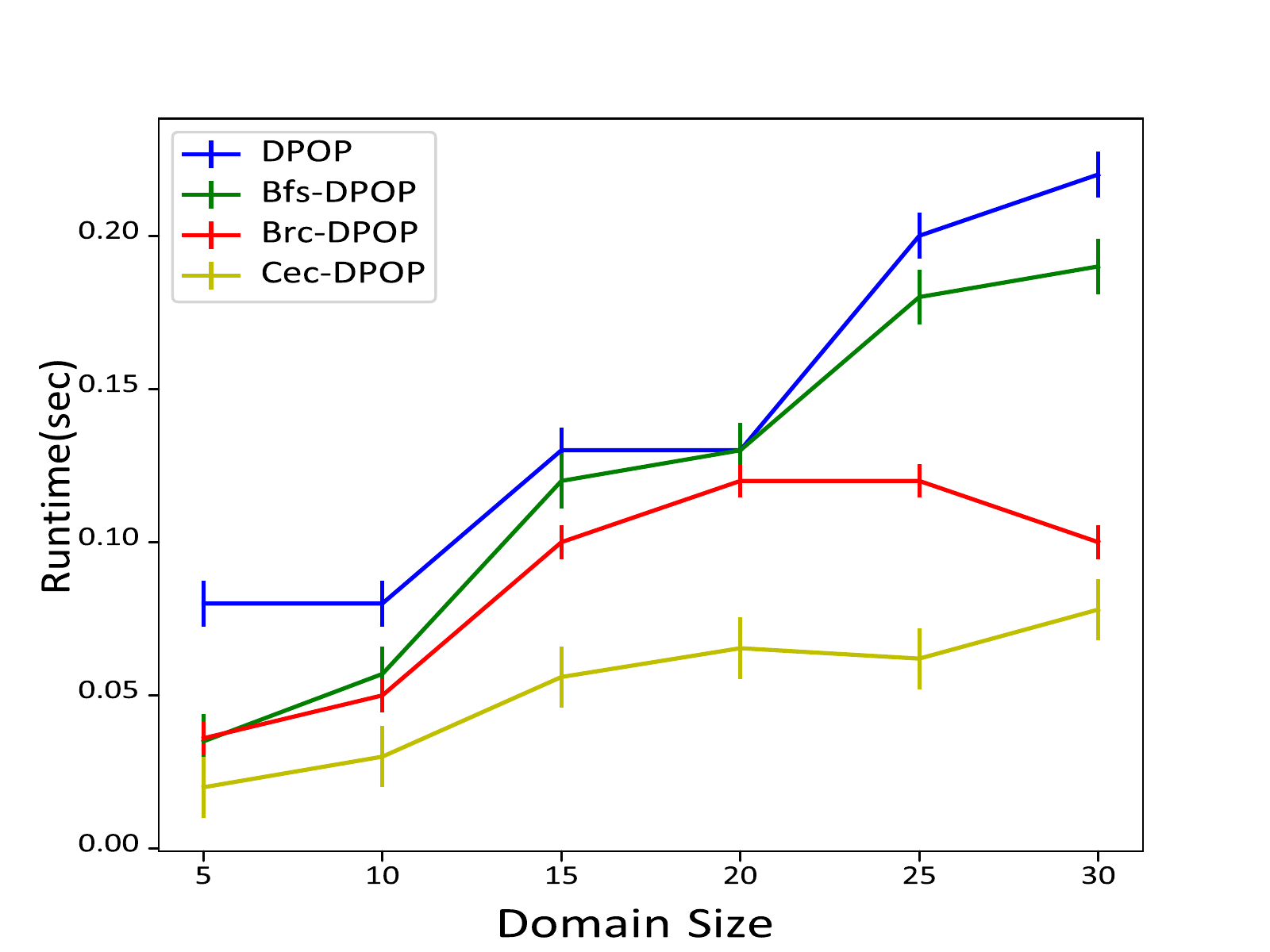}
			\caption{domain size vs runtime (random graph)}
			\label{graph:two}
		\end{subfigure}
		\hfill
		\begin{subfigure}[H]{.45\textwidth}
			\centering
			\includegraphics[scale=0.45]{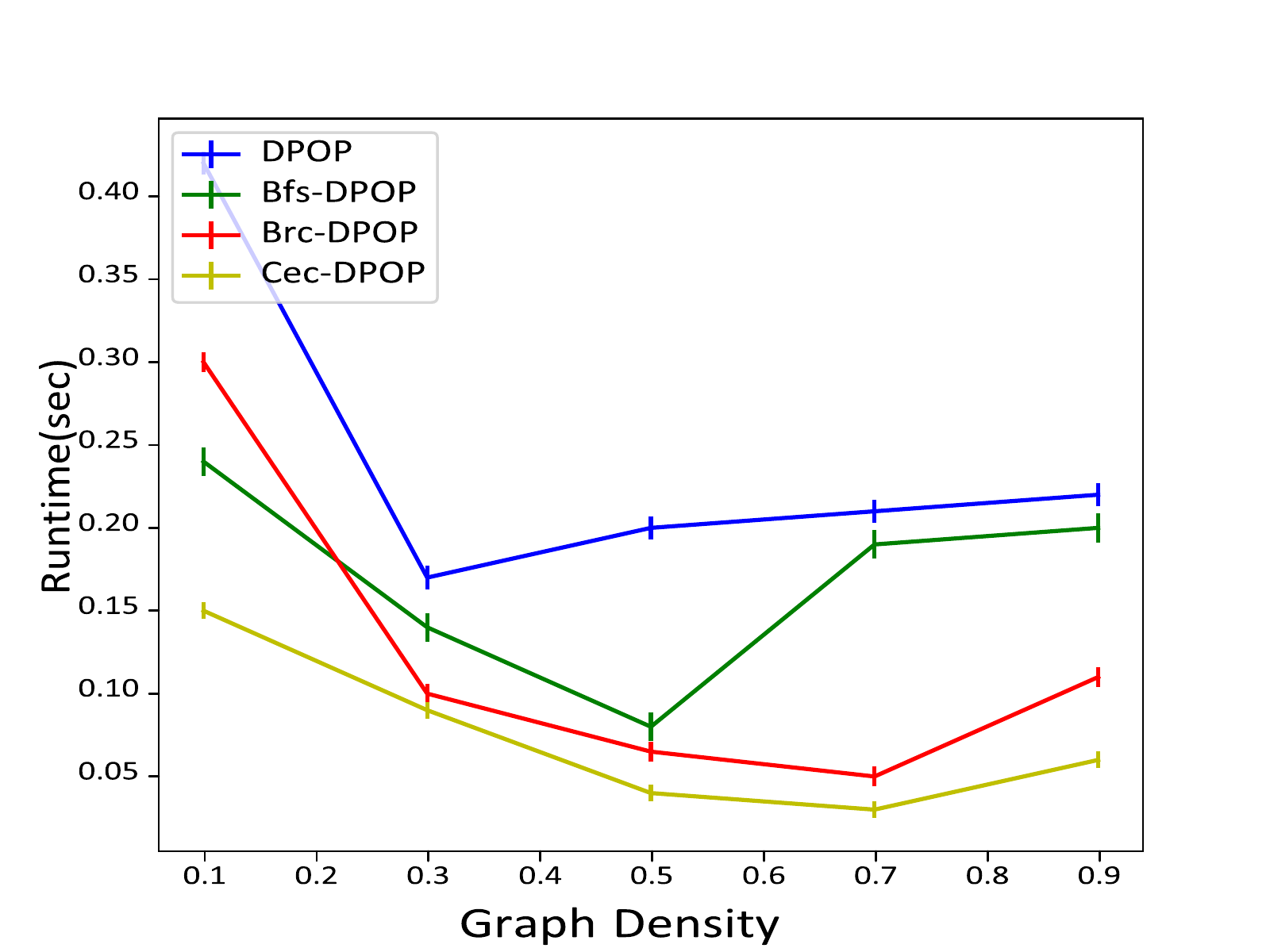}
			\caption{graph density vs runtime (random graph)}
			\label{graph:three}
		\end{subfigure}
		\hfill
		\begin{subfigure}[H]{.45\textwidth}
			\centering
			\includegraphics[scale=0.45]{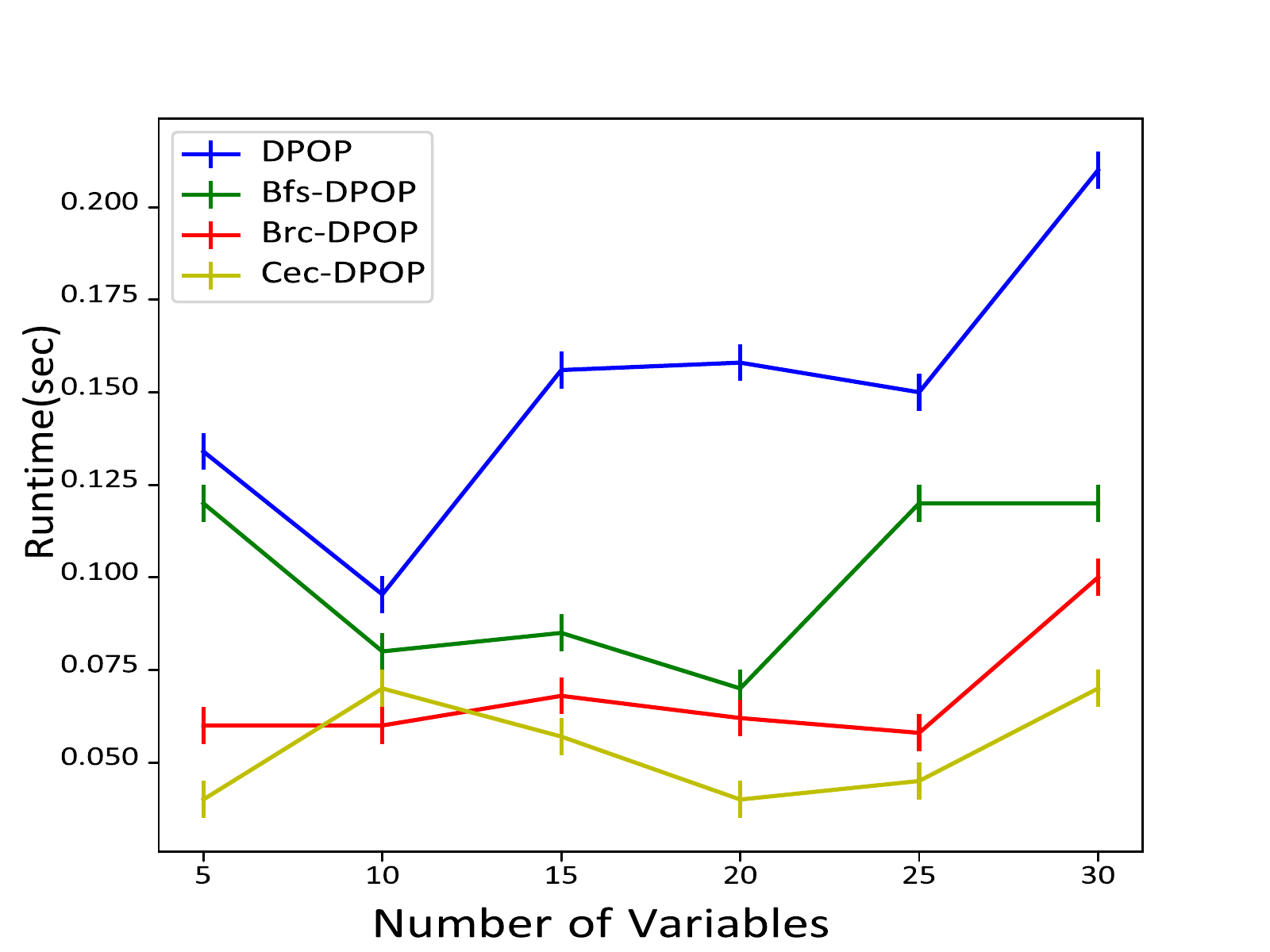}
			\caption{number of variables vs runtime (distributed RLFA problem)}
			\label{graph:rlfa}
		\end{subfigure}
		
		\caption{Experimental results for random graphs and the distributed RLFA problem}
	\end{figure*}

\section{Complexity Analysis}
\vspace{-2mm}
For enforcing cross-edge consistency we have constructed a path and next enforced arc-consistency. The path construction phase needs to send a message containing information about its subtree to its parent for each endpoint of the cross-edge. Therefore, the complexity of this phase is $O(|CE|\log(|X|))$, where $|CE|$ is the number of cross-edge and $|X|$ is the number of variables. The path construction phase requires the lowest common ancestor for each pair of nodes, which is found in a preprocessing phase having a complexity of $O(\log(|X|))$. The next phase is the arc-consistency enforcement phase. In this phase, each hard constraints are evaluated to check whether the domain of both variables connecting the endpoints is consistent with each other. Given the number of hard constraints is $C_H$ and the average domain of each variable is $d$, the complexity of this phase is $O(C_{H}d^{3})$. Here, in order to check whether each value in the domain of an endpoint is consistent with every value of the other endpoint, it  requires three nested loops thus resulting in $d^{3}$ computations. The final phase is enforcing cross-edge consistency. In this phase, each agent waits for its parent agent to send a CeC message which it uses to find the final cross-edge consistent matrix which requires a complexity of $O(d^{3})$ which is required for the multiplication of two matrices. Apart from regular matrix multiplication, entry-wise matrix multiplication has a complexity of $O(d^{2})$. The process continues for each variable, and as such the total complexity of cross-edge consistency enforcement phase is $O(|X|(d^{3} + d^{2}))$.

The arc-consistency enforcement phase requires $O(d|X|)$ messages, where the size of each message is $O(d)$. In each step of arc-consistency enforcement, the domain information of a variable is only required to be propagated. Therefore, the cross-edge consistency enforcement phase requires $C_{H}$ messages and the size of each message is $O(d^{2})$. In this phase, we only propagate CeC messages which contain the consistency matrices and the size of a message depends on the size of these matrices.

\section{Experimental Results}
We now empirically evaluate how much performance improvement can be attained using CeC-DPOP in comparison to the original DPOP algorithm and two important variants of DPOP named BFS-DPOP and BrC-DPOP. Unlike CeC-DPOP, the original DPOP uses DFS pseudo-tree as the communication structure and do not actively exploit hard constraints. Therefore, it is reasonable to observe the  attributes of CeC-DPOP (i.e. inclusion of soft constraints along with hard constraints and use of BFS pseudo-tree as the communication structure) with respect to the original DPOP. Additionally, We consider with BFS-DPOP algorithm as a benchmark because it also uses  BFS pseudo-tree as the communication structure. Finally, we compare CeC-DPOP with BrC-DPOP as both the algorithms can deal with DCOPs having both type of constraints. Note that another DPOP variants H-DPOP has not been considered as a benchmark because it already shown in the work of BrC-DPOP that it is outperformed for its high runtime. To benchmark the runtime of our algorithm CeC-DPOP as well as the benchmarks, we run our experiments on two different types of DCOP settings: random constraint graph and the distributed RLFA problems. These choices are made following the experimental settings of BrC-DPOP.

In case of random DCOPs, the runtime of the algorithms have been reported varying three parameters: number of variables, their domain size and graph density (i.e. the ratio of the constraint number and $n*(n-1)/2$ where n is the number of varaibles). For the first parameter, we vary the number of nodes from $5$ to $30$ in Figure ~\ref{graph:one} setting the domain size, $|D| = 10$ and edges are created by taking pairs of variables randomly and connecting them considering fix graph density, $\rho = 0.5$. For the second parameter, we execute the algorithms by changing domain size from 5 to 30 in Figure ~\ref{graph:two} where we consider the parameters number of variables $X$ at $20$ and graph density $\rho$ at $0.5$. Then for the third parameter, we increase the graph density from 0.1 to 0.9 in Figure ~\ref{graph:three} setting the number of variables and domain size as above mentioned. In case of Distributed RLFA Problem, we observe runtime by varying number of nodes from $5$ to $30$ in Figure ~\ref{graph:rlfa} and setting other parameters with the same as the previous setting. In both type of settings, we generated 30 instances and calculated the average runtime that we found by running each of the algorithms. All of the experiments were performed on a simulator implemented in an Intel \textit{i7} Octacore $3.4$GHz machine with 16GB of RAM. 

Our experimental results for solving random DCOPs are depicted in Figures~\ref{graph:one} -- \ref{graph:three}. In so doing, we generate three synthesized graphs. Specifically, we use hard constraints that are either ``less than", ``greater than" or ``equal" alongside soft constraints for which we randomly generated utility values from the range $[0,100]$. In Figure~\ref{graph:one}, we observe that runtime of Cec-DPOP increases in a steady way with respect to other algorithms as we increase the number of nodes. The result can be understood by observing the fact that the larger the constraint graph, the greater the advantages in parallelism and communication efficiency are found by CeC-DPOP. Another notable advantage of CeC is that it can avoid performing operations on the values pruned during the consistency enforcement phase. As a result, our algorithm is slightly faster than both BFS-DPOP and BrC-DPOP. More precisely, we observer that CeC-DPOP experiences $17 - 81\%$ reduction in runtime compared to DPOP, $25-62\%$ compared to Bfs-DPOP and $36-66\%$ compared to Brc-DPOP.

Figure ~\ref{graph:two} illustrates the results based on the next setting that is varying the domain size while setting the number of nodes and graph density, we observe that runtime of  Cec-DPOP increases at smaller rate than that of other algorithms. To be precise, we find $49 - 75\%$ smaller runtime than DPOP, $42-65\%$ than Bfs-DPOP and $22-48\%$ than Brc-DPOP. The reason behind this performance is that when domain size increases, more values in each domain is pruned by CeC-DPOP through consistency enforcement producing UTIL message of smaller dimension. As a result, the required time to compute messages decreases at a significant rate. Though Brc-DPOP have relatively smaller runtime than DPOP and BFS-DPOP for enforcing branch consistency, Cec-DPOP is always the winner through enforcing cross edge consistency.

In the third experimental setting, we vary the the graph density and set the other two parameters (Figure~\ref{graph:three}). we observe a notable performance gain of CeC-DPOP in terms of  runtime compared to other algorithms. To be exact, we detect $47 - 85\%$, $35-84\%$ and $10-50\%$ reduction of runtime with comparison to DPOP, Bfs-DPOP and  Brc-DPOP, respectively. This behavior can be explained when we notice that CeC-DPOP uses BFS pseudo-tree as the communication structure which is generated from dense constraint graph resulting more branches. As a result, more parallelism is experienced. Another reason is that the number of edges is relatively higher in the dense constraint graphs creating the opportunity of cross edge consistency enforcement at a significant level. Thus, more domain values are pruned and more shorter messages are produces resulting in a smaller computation time. Overall, a great reduction in runtime is observed. Another important behavior that should be mentioned is phase transition occurs for DPOP, BFS-DPOP and BrC-DPOP in Figure~\ref{graph:one} when number of variables goes across 20. DPOP and BFS-DPOP experiences phase transition in Figure~\ref{graph:two} when domain size increase towards 20. In both cases, transition for CeC-DPOP is relatively smooth than other algorithms.

As aforementioned, Distributed RLFA Problem is considered as the second type of problem to evaluate CeC-DPOP against the benchmarking algorithms. The distributed RLFA problem \cite{cabon1999radio} consists of a set of channels, each having a transmitter and receiver at both ends. The aim is to assign a frequency from a given set $F$ by minimizing the total interference at the receivers below an acceptable level and at the same time using as few and also as low frequencies as possible. For our experiment, we mapped a transmitter as a variable and for simplicity, we assigned a single agent to a variable. The domain of a variable consists of  frequencies (chosen from available spectral resources) that can be assigned to a variable. The interference between transmitter is modeled as a constraint of the form $x_i - x_j > s$ where $x_i$, $x_j$ are variables and $s$ is a random frequency separation. We varied the number of variables in Figure~\ref{graph:rlfa} and observed the runtime with DPOP, BFS-DPOP, and BrC-DPOP. We set the domain size, $|D| = 10$, $s\in \{3, 4\}$ and graph density, $\rho = 0.5$. Significantly, the results are similar as observed to that in case of random DCOPs. More specifically, CeC-DPOP outperforms other algorithms contributing $27 - 75\%$, $12-66\%$ and $16-35\%$ reduction in runtime relative to DPOP, Bfs-DPOP and  Brc-DPOP. It is worth mentioning that in the distributed RLFA problem and also in random DCOPs that was previously described, CeC-DPOP reduces message size around $5\%$ than Brc-DPOP.


\section{Conclusion}
    We present a new algorithm, CeC-DPOP, that significantly reduces the runtime of the DPOP algorithm that can be used to solve DCOPs having both soft and hard constraints. We empirically observe that our algorithm performs around 10-85\% faster than the state of the art algorithms. This is mainly possible because of the introduction of cross edge consistency. In addition, the use of the BFS pseudo tree as a communication structure enables CeC-DPOP to perform even faster. As a result, CeC-DPOP extends the use of DPOP in solving real-life problems that include both hard and soft constraints. In future work, we intend to investigate whether our approach can be applied to other DPOP extensions, as well as how much speedup can be achieved for them.

\bibliographystyle{ieeetr}
\bibliography{references}

\end{document}